\documentclass[a4paper,12pt]{article}

\usepackage[margin=3cm,footskip=1cm]{geometry}
\usepackage[T1]{fontenc}
\usepackage[utf8]{inputenc}
\usepackage[raggedright]{titlesec}
\usepackage{amsfonts,amsmath,bm,amssymb,amsxtra,amsthm,bm,mathtools,xspace,booktabs,url,lscape,graphicx,etoolbox,soul,color,multirow,subfigure,authblk,siunitx,lipsum,hyperref,fixme,float,latexsym}
\usepackage{eucal,mathrsfs,rotating,srcltx}

\numberwithin{equation}{section}
\theoremstyle{plain} 

\theoremstyle{definition} 

\makeatletter
\newcommand\CorrespondingAuthor[1]{%
  \begingroup%
  \def\@makefnmark{}%
  \footnotetext{Corresponding author: #1}%
  \endgroup%
}
\makeatother
\makeatletter
\renewenvironment{abstract}{%
  \small%
  \begin{center}%
    \bfseries \abstractname\vspace{-.5em}\vspace{\z@}%
  \end{center}%
  \quote%
}{\endquote}
\makeatother
\makeatletter
\DeclareRobustCommand*\subref{\@ifstar\sf@@subref\sf@subref}
\makeatother

\begin{document}

\title{
Estimating the three-month series of the Chilean Gross Domestic Product
}

\author[1]{Christian Caama\~no Carrillo}
\affil[1]{Departamento de Estad\'istica, Universidad del B\'io-B\'io, Concepci\'on, Chile,  \texttt{chcaaman@ubiobio.cl}}

\author[2]{Sergio Contreras}
\affil[2]{Departamento de Estad\'istica, Universidad del B\'io-B\'io, Concepci\'on, Chile,  \texttt{scontre@ubiobio.cl}}

\date{\today}

\maketitle

\begin{abstract}
In this paper the methodology proponed by \cite{CPF}; is applied to estimate the three-month series of the Chilean Gross Domestic Product (GDP) in the period composed between 1965 and 2009. In first place, the equation of Engle-Granger is estimated using the data of the yearly GPD and related variables. The estimated coefficients of this regression are used to obtain a first estimation of the three-month GDP with measurements errors. Then a State Space model is estimated through Benchmarking in order to improve the preliminary estimation of the GDP with the purpose of eliminating the maximum error of measurement and in order that the sum of the three-month values coincide with the yearly GDP.\\

\noindent\textit{Keywords:}
Benchmarking; Engle-Granger equation; Kalman Filter (KF); State Space Models; GDP
\end{abstract}

\section{Introduction}\label{int}

An important limitation in order to specify and estimate a macroeconomic model that describes the Chilean economy is to use variables with sufficient number of observations that allow a reliable econometric estimation. Among these variables, the gross domestic product (GDP) constitutes a fundamental magnitude. However, for this variable in Chile there is no quarterly information before 1986 according to the new splice methodology. Therefore, although annual information is available for this series, quarterly information is only available as of 1986.

The methodology proposed by \cite{CPF} allows solving situations like the one described above. They apply their methodology to GDP, but it can be considered for other series.

GDP is one of the most relevant economic indicators when measuring the economic power of a country. It constitutes the broadest measure of economic activity, containing a wide range of constitutive factors representing different areas of the economic activity. We can define it, as the total value of products and services produced by a nation in a specific period of time, usually a year or a quarter.

The last assessment is of particular interest in this article, that is, the frequency with which GDP is calculated. For various economic studies, a low number of data can cause serious problems in terms of the quality of a quantitative analysis. Due to the above, it would be useful to make quarterly estimates of GDP values from the official data.

Therefore, the main objective of this article will be to describe and apply a methodology for the estimation of Chilean quarterly GDP from 1965 to 2009. The procedure consists of an initial estimate of the quarterly GDP, which is obtained through the static Engle-Granger equation, (see \cite{EG1} and \cite{EG2}) or cointegration ana\-lysis using the annual GDP data and variables related to GDP. The estimated coe\-fficients of this regression are used to construct a quarterly equation between GDP and related variables, by interpolating the estimated coefficients with the quarterly data of the aforementioned variables. This equation produces a first estimate of the quarterly GDP, called dirty GDP, due to the presence of errors.

The second stage is to improve the estimate by minimizing measurement errors and considering that this measurement must be consistent with the annual GDP calculated by the Central Bank of Chile. This is, the sum of the quarterly estimates must be equal to the total annual.

The process of harmonization of quarterly and annual estimates is known in the literature as Benchmarking and in this article is addressed using the State Space  approach. See \cite{DK} and \cite{H}.

The article is organized as follows: the next section briefly analyzes the methods used for estimation. In the next stage, the methodology of quarterly GDP estimation based on the proposal of \cite{CPF} is described. Then, the results of the estimated models are introduced and discussed. Finally, the conclusions are presented.

\section{Methods}\label{met}

\subsection{Engle-Granger Cointegration Analysis}\label{ega}

Cointegration analysis is a frequently used technique  in the study of time series. The concept of cointegration was introduced by Granger and subsequently studied in depth in \cite{EG1} and \cite{EG2}. This technique emerges as a procedure that allows to discriminate between real relationships and spurious relationships between variables. A spurious regression arises when we try to relate two variables between which there is no type of cause-effect relationship through a regression and it is erroneously concluded, after regression, that such a relationship exists. According to what is stated in \cite{GN}, one of the characteristics of spurious relations consists in having a very high coefficient of determination and a Durbin-Watson statistic close to zero.

From an econometric point of view, two or more time series that are of order $I(1)$ are cointegrated if there is a linear combination of these series that is stationary or of order $I(0)$.

From an economic point of view, cointegration can be seen as a long-term equilibrium relationship between variables, such that these variables can deviate from the equilibrium situation in the short term, but with time they will return to the equilibrium. Another definition from the economic point of view says that two or more series are cointegrated if they move together over time and the differences between them are stable i.e, stationary, even when each series is not stationary. The differences (or error term) in the cointegration equation are interpreted as the imbalance error for each particular point of time.

\subsection{Benchmarking method}\label{bm}

The concept of Benchmarking can be seen as a special form of signal extraction, which occurs when two (or more) data sources are available for the same target variable, which have different temporal frequencies, for example, monthly vs. annual, monthly vs. quarterly, annual vs. quarterly, etc. Genera\-lly the two data sources do not correspond, for example the quarterly sums of the measurements of a variable are not equal to the corresponding annual measure. Moreover, one of the data sources is more accurate than the other, it is usually the least frequent because it originates, for example, in censuses, hypothetically assumed free of sampling errors. The most reliable source is considered a Benchmark (comparative framework).

More specifically, Benchmarking is the process of trying to fit the most frequent series to Benchmark. This is done by decomposing the series into its structural elements: trend, seasonality, cycle, irregularities and measurement errors, where the sum of these elements excludes the error.

According to \cite{CPF}, there are two main methods for applying benchmarking to a time series: a purely numerical approach and a statistical method. the numerical approach covers the family of methods based on the minimization of a squared sum proposed by \cite{D}. the statistical method, in turn, may be based on ARIMA processes such as those described by \cite{HT}; the state space models proposed by \cite{DQ}, or the models that use a group of regressions, such as \cite{CD}.

\section{Methodological procedure}\label{promet}

\subsection{First Stage: Analysis of Cointegration and Obtaining Dirty GDP}\label{etap1}

An initial estimate of the quarterly GDP is obtained, which is obtained through the static Engle-Granger equation or cointegration analysis using the data of annual GDP and variables related to GDP. Then interpolation is applied using the coe\-fficients estimated by the previous equation with the quarterly data of the related variables. Thus obtaining what is known as dirty GDP. The variables to be used are: the monetary aggregate, the price of copper, the terms of trade, the exports of goods and services and the mi\-ning production index obtained from the page of the Central Bank of Chile, on an annual and quarterly basis. Specifically, the calculation of dirty quarterly GDP is based on the estimation of a regression of GDP against the aforementioned series, with annual frequency expressed in base 1986 indices in order to obtain long-term coefficients that relate to the variables. After having estimated the cointegration vector by OLS, we form a linear combination using the quarterly frequency series in order to obtain the interpolated series of the quarterly GDP. The estimated quarterly series are linked and compared with the series of the Central Bank of Chile, producing the dirty GDP series, which is perfected in the second half of the procedure.

Prior to 1986, GDP data were only available on the annual frequency, which explains why the estimate was made on this frequency. It is of great importance that the estimated dirty GDP series incorporate the components present in the original series of GDP obtained by the Central Bank of Chile after 1986 and respect the annual information between 1965 and 2009 of the original series of GDP.

Therefore, dirty GDP is the first approximation of the quarterly GDP that recovers the lack of data prior to 1986 and that have a seasonal pattern thereafter identical to the series of the Central Bank of Chile.

\subsection{Second Stage: State Space Models by Benchmarking and obtaining clean GDP}\label{etap2}

The State Space approach provides a unified methodology for studying a wide va\-riety of problems in time series. For example, by means of this approach it is possible to model the behavior of the different components of a series separately, and immediately combine these sub-models in obtaining a single model for the series of interest. In this context, the State Space models are called Structural Models. See for example, \cite{DK} and \cite{H}. The State Space models are formed by two types of variables: the non-observable variables of the state, which determine the movement of the system in time, and the observations of the series.

In general situations, the Gaussian general model of state space is defined by:
\begin{eqnarray}\label{ec1}
y_t  &=&  Z_t\alpha_t+\epsilon_t, \; \epsilon_t \sim
N(0,H_t), \nonumber \\
  \alpha_{t+1} &=&  T_t\alpha_{t}+R_t\eta_t,\;
\eta_t \sim N(0,Q_t),\; t=1,\ldots,n,  \nonumber\\
\mbox{with}&\;&\alpha_1\sim N(a_1,P_1)
\end{eqnarray}
where $y_t$ is a $p\times 1$ vector of observations; $\alpha_t$ is an unobserved vector of dimension $m\times 1$ called state vector and the terms of independent perturbations, $\epsilon_t$ and $\eta_t$ are serially independent assumptions independent of each other at all times.

Matrices $Z_t, T_t, R_t, H_t$ and $Q_t$, with the appropriate dimensions, are cons\-tants and they called arrays of the system, assumed as known. In practice some or all of these matrices depend on a vector of unknown hyperparameters $\psi$, whose estimation is described, for example, in \cite{DK} and \cite{H}. It is assumed that $R_t$ is a subset of columns of $I_m$; $R_t$ is called the selection matrix, which chooses the lines of the state equation that have non-null perturbations.

Estimation of model (\ref{ec1}) is done by Kalman filter in combination with maximum likelihood. The Kalman filter is a recursive algorithm that allows to update the knowledge of the system every time a new observation arrives, and allows to calculate the optimal estimator of the state vector based on the available information up to the time $t$. It can be obtained through the application of existing results in the multivariate regression theory with Gaussian perturbations, see \cite{DK} and \cite{H}. Some applications, using Kalman Filter can be found in \cite{A} and \cite{ABK}.

State Space Models by Benchmarking have been studied to some extent by \cite{DQ}. One of the corresponding formulations in the form of State Space proposed by \cite{DQ} was revised, later by \cite{DK}.

This article presents a probabilistic version equivalent to model \cite{DQ}. It takes $y_t$ as dirty GDP (theorically with measurement errors), $x_t$ like the dirty GDP of the Central Bank of Chile (theorically free of measurement errors). The state vector, according to the formulation of \cite{DQ} is given by:
$$ \alpha_t= \left[  \mu_t,\ldots, \mu_{t-3}, \gamma_t, \ldots,\gamma_{t-3},\epsilon_t,\ldots,\epsilon_{t-3}, u_t\right]'$$
where $\mu_t$ is local level, $\gamma_t$ is the stochastic seasonal component, $\epsilon_t$ is the irregular component and $u_t$ is the measurement error associated with dirty GDP, which assumes a stationary model $AR(1)$ with unit variance.

The matrix $Z_t$ is defined as
$$Z_t=\left[
  \begin{array}{ccccccccccccc}
    1 & 0 & 0 & 0 & 1 & 0 & 0 & 0 & 1 & 0 & 0 & 0 & 1 \\
  \end{array}
\right]$$
if $t$ is not a multiple of four, and
$$ Z_t = \left[ \begin{tabular}{ccccccccccccc}

           1 & 0 & 0 & 0 & 1 & 0& 0& 0 & 1 & 0 & 0 & 0 & 1 \\
           1 & 1 & 1 & 1 & 1 & 1 & 1 & 1 & 1 & 1 & 1 & 1 & 1 \\

         \end{tabular} \right]
$$
if $t$ is multiple of four.

In the other hand, matrix $H_t$ is null for every $t$. The other matrix associated to the state vector are invariant in time. Note that, given the formulation of the State Space, the coherence between clean GDP (defined as $y_t^*=y_t-u_t$) and annual GDP, the Central Bank of Chile always occurs in multiples of four periods.

\section{Results}\label{res}

\subsection{Dirty GDP obtention}\label{res1}

The results of the initial estimate of the quarterly GDP, obtained through the static Engle-Granger equation, associated with the first stage of the methodology, are reported in table \ref{tab1}. The series of GDP of the Central Bank of Chile expressed in the Index with annual frequency with base 1986 corresponds to the dependent variable. The independent variables co\-rrespond to a linear trend term (TT), the logarithm of the monetary aggregate (L(AM)), the price of copper (PC), the logarithm of the terms of trade (L(TIR)), the logarithm of exports of goods and services (L(EBS)) and the mining production index (MPI). The adjusted coefficient of determination $R^2$ and the Durbin Watson statistic are also reported.

\begin{table}[htbp]
\begin{center}
\scalebox{0.8}{
\label{tab1}
\begin{tabular}{ccccc}
  \hline
  Variable & Coef. & Std. Error & t$-$Stat. & prob. \\ \hline
  TT & 0.0335 & 0.0028 & 11.797 & 0.00 \\
  L(AM) & -0.0829 & 0.0083 & -9.9362 & 0.00 \\
  PC & 0.0790& 0.0137& 5.7490 & 0.00 \\
  L(TIR) & -0.3395 & 0.0539 & -6.2986 & 0.00 \\
  L(EBS) & 0.5950 & 0.0612 & 9,7139 & 0.00 \\
  IPM & 0.2397 & 0.0568 & 4.2145 & 0.00 \\ \hline
  R$-$squared & 0.9942 & Mean dep. var &  & 1.54\\
  Ad. R$-$squared & 0.9935 & S.D. dep. var & & 0.88\\
  S.E. of reg. & 0.0712& AIC &  & -2.32 \\
  Sum sq. resid & 0.1960 & BIC &  & -2.08\\
  Log Lik.&  58.228& DW &  & 1.76\\
  \hline
\end{tabular}
}
\end{center}
\caption{Results of the Engle-Granger procedure between annual GDP and variables related to GDP.}\label{tab1}
\end{table}

It is clearly observed in table \ref{tab1} that the coefficients are significant and the signs are those expected for the related variables. On the other hand, there is a high coefficient of determination that indicates the good fit of the model, in addition to a Durbin Watson statistic that indicates a independence in the residues, making it possible to ensure that there is no spurious regression.

The unit root test of Dickey-Fuller increased for the residues of the model, which is presented in Table \ref{tab2}.
\begin{table}[htbp]
\begin{center}
\scalebox{0.8}{
\label{tab2}
\centering
\begin{tabular}{l|l|l}
\hline
 \multicolumn{3}{c}{Residuals}\\\hline
                         Trend    & Intercept & None  \\\hline
$5.76 $ & $5.83$ & $5.91$\\
  \hline
\end{tabular}
}
\end{center}
\caption{Augmented Dickey-Fuller test (ADF) for residuals.}\label{tab2}
\end{table}

From table \ref{tab2}, we reject the null hypothesis of non-cointegration and conclude that the residues are cointegrated of order $I(0)$.

\begin{figure}[htbp]
\centering
\includegraphics[width=8cm, height=5cm]{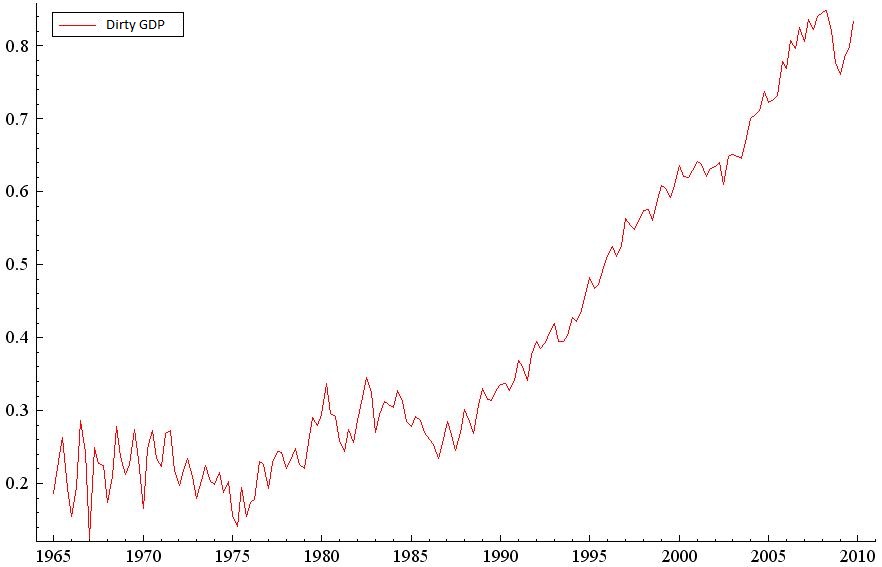}
\caption{Quarterly dirty GDP.}
\label{fig1}
\end{figure}

By interpolating the estimated coefficients of table \ref{tab1} with the quarterly series of the related variables, we obtain the estimated series of GDP. It is related to the series of quarterly GDP of the Central Bank of Chile from 1986 and is considered as the first approximation of the Chilean quarterly GDP, called dirty GDP. Figure \ref{fig1} shows the dirty GDP series.
Clearly, the estimated dirty GDP incorporates the components present in the quarterly series of Chilean GDP after 1986 as well as the cyclical dynamics of that period. It also and respects  the annual information between 1965 and 2009 of the original series of GDP.
\subsection{Clean GDP obtention}\label{res2}

It is preceded to estimate the clean GDP by the model of State Space by Benchmarking. To avoid the heteroscedasticity problem, a dirty GDP transformation is performed assuming that the variance of the error is proportional to the square of the export of goods and services (EBS) thus successfully capturing the inconsistency of the variance.

Using the programming language Ox, the clean GDP estimation is performed, with the STAMP package of \cite{KHDS} and the SsfPack package of \cite{KSD} that use the approximation BFGS method for the maximization process.

Table \ref{tab3} and Figure \ref{fig3} show the statistical summary of the estimated variances and the components referring to the estimation of the State Space by Benchmarking model for the normalized dirty GDP series, in addition to the verification of model assumptions.

\begin{table}[!hbtp]
\begin{center}
\scalebox{0.8}{
\label{tab3}
\begin{tabular}{c|c}
\hline
Log-Likelihood& 288.559\\ \hline
Level Variance $\sigma_\varepsilon^2$ & 0.00053860\\\hline
Seasonality variance $\sigma_w^2$	 & 0,00011113\\\hline
Irregularity variance $\sigma_\epsilon^e$& 	0,0075303\\\hline
Measurement Error Variance $\sigma_u^2$ &	0,00047772\\\hline
MSE &	0,0097188\\\hline
Pseudo $ R^2$	& 0,966332\\\hline
AIC &	-0,532673\\ \hline
\end{tabular}
}
\end{center}
\caption{Information about the estimated model of State Space by Benchmarking.}\label{tab3}
\end{table}

\begin{figure}[!hbtp]
\centering
\includegraphics[width=8cm, height=7cm]{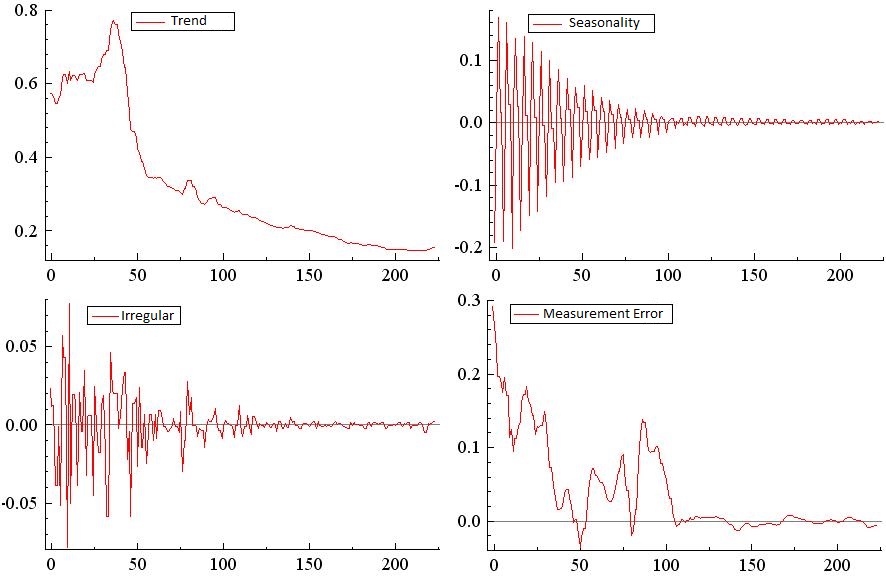}
\caption{Estimations of the components of the model.}
\label{fig2}
\end{figure}

The figure \ref{fig2}, give us the components of the estimated model of the normalized GDP series.

Changes are confirmed over time in figure \ref{fig2}, indicating movements of both growth and decrease for certain periods of time (see trend). On the other hand, it can be seen that the peaks relative to the quarters do not remain constant over time (see seasonality).

\subsection{Standardized Residuals Analysis}\label{res3}

The residuals of the model must be independent, with zero mean and unit variance, where the considered tests can be graphical, such as QQ-plot, correlogram and histogram; and tests of specification among which are, Durbin-Watson, Ljung-Box and Jarque Bera.

\begin{figure}[htbp]
\centering
\includegraphics[width=8cm, height=7cm]{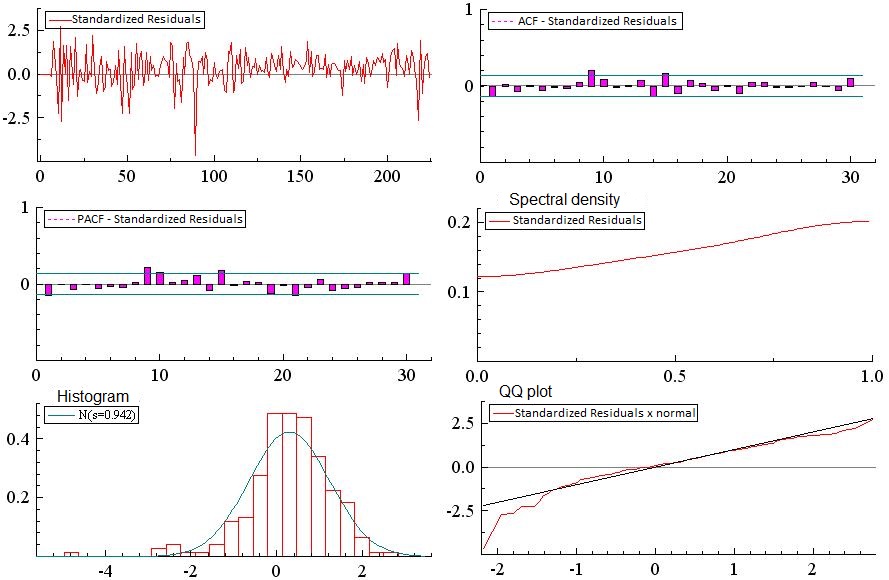}
\caption{Standarized Residuals analysis.}
\label{fig3}
\end{figure}

Figure \ref{fig3} shows the correlograms, QQ-plot and histograms of the residuals of the model.

Clearly, by observing the autocorrelation function, it can be concluded that the residuals are independent; and on the other hand there is no normality in the resi\-duals. The following table contrasts the evidence previously presented with more formal statistical tests.

\begin{table}[!hbtp]
\begin{center}
\scalebox{0.8}{
\label{tab4}
\begin{tabular}{c|c}
\hline
Ljung-Box Test  &	5,9823   (0,2005) \\ \hline
Jarque-Bera-Test	& 154,325 (2,2e-16) \\ \hline
Dickey-Fuller Test& 	-7,2319  (0,01) \\ \hline
F Test of Heteroscedasticity & 	0,58423 \\ \hline
\end{tabular}
}
\end{center}
\caption{Diagnostic test of residuals standardized.}\label{tab4}
\end{table}

When observing the statistical tests, it can be concluded that the assumptions of homoscedasticity, independence and stationarity of the residuals in the estimated model are met.

\subsection{Clean GDP}\label{res4}

Finally, in order to obtain the estimated clean GDP, we must multiply the estimation of the model of State Space by Benchmarking with the macroeconomic variable that standardized dirty GDP. This is, the export of goods and services (EBS).

\begin{figure}[htbp]
\centering
\includegraphics[width=8cm, height=5cm]{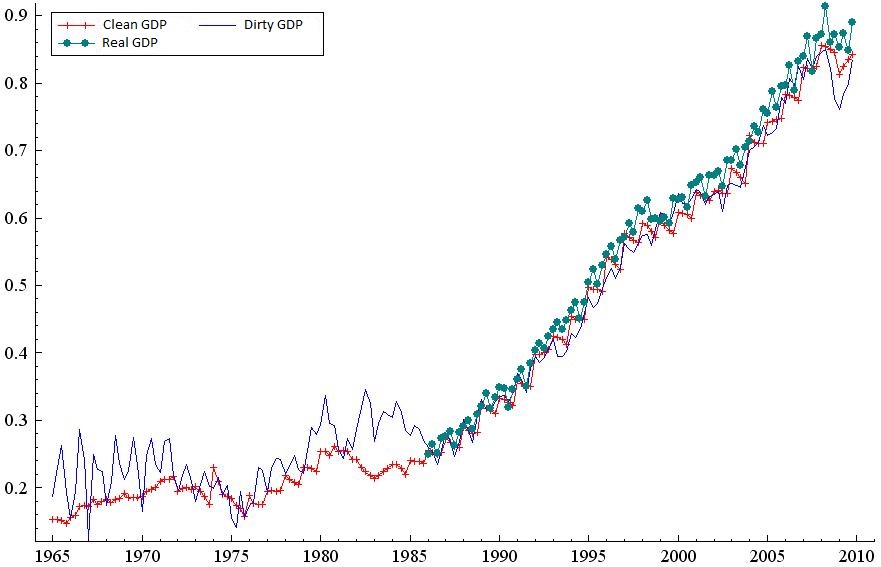}
\caption{Dirty and clean GDP.}
\label{fig4}
\end{figure}

Figure \ref{fig4} presents a graph that compares clean GDP with dirty GDP and quarterly GDP calculated by the Central Bank of Chile after 1986.

From figure \ref{fig4}, it is clearly observed that dirty quarterly GDP presents an overestimation between the period (1981-1986) that does not represent the annual GDP characteristic calculated by the Central Bank of Chile. This allows the proposed procedure to correct the quarterly operation and the measurement error by harmonizing the annual and quarterly GDP estimates. Thus, clean GDP incorporates the seasonal component in the quarterly series and respects the annual information of the series submitted by the Central Bank of Chile.

\section{Conclusion}

This article was based mainly on: (1) the methodology for
estimating the Chilean quarterly GDP between the 1965-2009
periods, (2) demonstrating the empirical methodology, and (3)
reporting and analyzing the resulting series. The
importance of quarterly GDP estimation for periods
before 1986 arises from the need to cover in empirical studies
a longer period of the Chilean economy is history and the need
to use variables with a sufficient number of observations that
allow a reliable econometric estimation.

The results have shown a fairly satisfactory relationship with
the diagnosis of residuals. The cleaning of the estimated GDP
with State Space models by Benchmarking corrects the
measurement error with respect to dirty GDP. Therefore, the
estimate is consistent with the annual GDP and provides an
attractive estimation method.

There are other studies that present two directions. The first of
these consists in applying Benchmarking to multivariate
models, which theoretically give greater accuracy and
reliability to the quarterly GDP. The second refers to using the
methodology to obtain the predicted quarterly GDP.

An estimate of the Chilean quarterly GPD could also be made, using benchmarking models for
the other methods mentioned, involving models based on ARIMA processes \cite{HT} and models that
use a group of regressions, such as \cite{CD}.

\section*{Aknlowledgements}
We are grateful to Internal Initiation Project DIUBB 1734082/I of the Universidad del B\'io-B\'io of Christian Caama\~no Applied Mathematical Research Group GI 172409/C of Universidad del B\'io-B\'io, Concepci\'on, Chile.







\end{document}